\documentclass[fleqn,usenatbib]{mnras}

\usepackage{newtxtext,newtxmath}

\usepackage[T1]{fontenc}

\DeclareRobustCommand{\VAN}[3]{#2}
\let\VANthebibliography\thebibliography
\def\thebibliography{\DeclareRobustCommand{\VAN}[3]{##3}\VANthebibliography}

\usepackage{graphicx}	
\usepackage{amsmath}	
\usepackage{ulem}

\newcommand{\kms}{\,km\,s$^{-1}$}
\newcommand{\msun}{\,M$_\odot$}
\newcommand{\vsini}{$V\sin i$}
\newcommand{\befrac}{$f_{\rm be}$}
\newcommand{\befracmax}{$f_{\rm be}^{\rm max}$}

\title[Rotation in YMCs]{Tracing stellar rotation in young massive LMC clusters}

\author[S. Kamann et al.]{%
Sebastian Kamann,$^{1}$\thanks{E-mail: s.kamann@ljmu.ac.uk (SKA)} 
Nate Bastian,$^{1,2,3}$\thanks{E-mail: nate.bastian@dipc.org} 
Florian Niederhofer,$^{4}$ 
Andrea Bellini,$^{5}$ 
Ivan Cabrera-Ziri,$^{6}$ \and 
Stefan Dreizler,$^{7}$
Fabian Göttgens,$^{7}$
Vera Kozhurina-Platais,$^8$
Mattia Libralato,$^9$
Sven Martens,$^7$ \and
Sara Saracino,$^{10}$ \\
\\
$^{1}$Astrophysics Research Institute, Liverpool John Moores University, IC2 Liverpool Science Park, 146 Brownlow Hill, Liverpool L3 5RF, UK\\
$^{2}$Donostia International Physics Center (DIPC), Paseo Manuel de Lardizabal, 4, 20018, Donostia-San Sebasti\'an, Guipuzkoa, Spain\\
$^{3}$IKERBASQUE, Basque Foundation for Science, 48013, Bilbao, Spain \\
$^4$Leibniz-Institut für Astrophysik Potsdam, An der Sternwarte 16, D-14482 Potsdam, Germany \\
$^5$Space Telescope Science Institute, 3700 San Martin Drive, Baltimore, MD 21218, USA \\
$^6$Vyoma GmbH, Karl-Theodor-Straße 55, 80803 Munich, Germany \\ 
$^7$Institut f\"ur Astrophysik und Geophysik, Georg-August-Universit\"at G\"ottingen, Friedrich-Hund-Platz 1, 37077 G\"ottingen, Germany\\
$^8$Eureka Scientific, Inc, 2452 Delmer St. Suite 100, Oakland, CA 94602-3017, USA\\
$^9$INAF - Osservatorio Astronomico di Padova, Vicolo dell'Osservatorio 5, Padova I-35122, Italy\\
$^{10}$INAF – Osservatorio Astrofisico di Arcetri, Largo E. Fermi 5, 50125 Firenze, Italy\\
}

\date{Accepted XXX. Received YYY; in original form ZZZ}

\pubyear{2025}

\begin{document}
\label{firstpage}
\pagerange{\pageref{firstpage}--\pageref{lastpage}}
\maketitle

\begin{abstract}
We present a detailed analysis of stellar rotation along the main sequences of NGC~1866 and NGC~1856, two young ($\sim$200-300~Myr) massive clusters in the Large Magellanic Cloud, using MUSE integral field spectroscopy. Differences in stellar rotation have been proposed as an explanation for the extended main sequence turn-offs and split main sequences in these clusters. In agreement with this idea, we find strong links between the photometric colours of the cluster stars and their projected rotation velocities, \vsini{}. While stars blueward of the split main sequences are characterized by a range of relatively low spins, those with redder colours are fast rotators. Following a statistical correction for inclination, we measure mean equatorial velocities for the red main-sequence stars in both clusters of $V_{\rm eq}=300$~\kms, corresponding to 70-80\% of the critical values predicted for such stars by current stellar models. We discuss these findings in the context of the different scenarios proposed to explain the stellar rotation distributions of young massive clusters. We further investigate whether the high rotation rates provide a natural explanation for the high fractions of Be stars we observe in both clusters, peaking at $\geq$50\% at the turn-off. We argue that if $\sim85\%$ of the critical rotation rate is high enough to trigger the formation of a decretion disk, most upper main sequence stars in the clusters are expected to become Be stars before leaving the main sequence.
\end{abstract}




\section{Introduction}

The Magellanic Clouds host rich populations of massive ($M_{\rm cl}\sim10^5\,{\rm M_\odot}$) star clusters that are indispensable probes for enhancing our understanding of stellar evolution. Contrary to the Milky Way, the Clouds have experienced periods of enhanced star formation through the last few giga-years \citep[e.g.,][]{nidever2020}. As a result, they host massive star clusters over a wide age range \citep[see][for a recent compilation]{horta2021}, including young massive clusters (YMCs, $t\lesssim500\,{\rm Myr}$) and intermediate-age massive clusters (IMCs, $500\,{\rm Myr}\lesssim t \lesssim 5\,{\rm Gyr}$) that are either absent in the Milky Way, or covered by many magnitudes of Galactic extinction. Crucially, the Clouds enable us to study the properties of massive star clusters as a function of age. They reveal how massive clusters evolve dynamically \citep[e.g.,][]{song2021} or trace the formation of multiple populations across cosmic time \citep[e.g.,][]{martocchia2018}.

A surprise has been the discovery of peculiar features in the colour-magnitude diagrams (CMDs) of YMCs and IMCs. \citet{mackey2007} discovered an extended main-sequence turn-off (eMSTO) in NGC~1846 ($t\sim1.5~{\rm Gyr}$), a feature that has subsequently been found to be ubiquitous in the CMDs of clusters younger than 2~Gyr, while those with ages $\lesssim$500~Myr additionally show split main sequences \citep[e.g.,][]{milone2015,milone2018}. Although differences in the stellar rotation rates of the cluster stars have been proposed as a plausible mechanism causing the features early on by \citet{bastian2009}, age spreads of 10s-100s of Myr within the clusters had been the most frequently discussed cause in the literature for numerous years \citep[e.g.,][]{goudfrooij2014}. However, \citet{niederhofer2015} found a strong correlation between the photometrically inferred age spreads of the clusters and their ages, which confirmed a key prediction of the stellar rotation scenario.

Indeed, the first spectroscopic measurements of eMSTO stars in NGC~1866 ($t\sim 200~{\rm Myr}$) by \citet{dupree2017} revealed a wide range in rotation rates. Shortly after, \citet{marino2018} found that the red main-sequence stars of NGC~1818 (40~Myr) to be rotating faster than their counterparts on the blue main sequence, while a similar signature was found among eMSTO stars in the SMC cluster NGC~419 ($t\sim1.5~{\rm Gyr}$) by \citet{kamann2018b}, where the reddest stars rotated the fastest. Using the high multiplexing capabilities of the MUSE integral field spectrograph, \citet{kamann2020} showed that the eMSTO of NGC~1846 hosts distinct populations of fast and slow rotators with mean projected rotational velocities (\vsini) of 60\kms and 140\kms, respectively. The latter, which make up 50-60\% of the bright stars, disappear at the bottom of the eMSTO, indicating the emergence of convective envelopes and magnetic braking in stars of masses $\lesssim$1.5\msun \citep{georgy2019}. The MUSE study of the 100~Myr old cluster NGC~1850 by \citet{kamann2023} highlighted a link between \vsini{} and colour across the eMSTO of the cluster and measured mean \vsini-values of 100\kms and 200\kms at either side of the cluster's split main sequence. While there is mounting evidence that stellar rotation is responsible for shaping the photometric features of these clusters, \citet{cordoni2022} showed that age spreads $\gtrsim$10~Myr can be ruled out in NGC~1818, based on its narrow main-sequence turn-\emph{on}. 

We highlight that while split main sequences and eMSTOs have initially been discovered in massive Magellanic Cloud clusters, the impact of stellar rotation can also be traced in low-mass open clusters of the Milky Way. Using \textit{Gaia} data, \citet{cordoni2018} reproduced the relation between cluster age and apparent age spreads from \citet{niederhofer2015} for Galactic clusters. \citet{bastian2018} found a link between colour and \vsini{} in the eMSTO of NGC~2818 ($t\sim800~{\rm Myr}$), which was subsequently also discovered in other low mass Galactic clusters \citep[e.g.,][]{lim2019,maurya2024,cordoni2024}.

The recent spectroscopic discoveries have raised the question about the origin of the stellar spin bimodality in young and intermediate-age clusters, with no consensus yet. Motivated by the finding of \citet{abt2004} that early-type stars in tight binary systems rotate slower than their single-star equivalents, a difference attributed to braking via binary tidal interactions, \citet{dantona2015} proposed that all stars were born as fast rotators, and that the slow rotators would result from binary interactions. Similarly, \citet{wang2022} also proposed that all stars initially had high angular momenta, but that the population of slowly rotating stars was formed via stellar mergers. Such mergers would also rejuvenate the stars on the blue main sequence, which often appear bluer (i.e. younger) than predicted by slow- or non-rotating stellar models \citep[e.g.,][]{dantona2017}. Finally, \citet{bastian2020} argued that the stellar spin of a star should depend on the lifetime of its circumstellar disk during the pre-main sequence phase, in the sense that stars losing their disks early can spin up to high rotation velocities, while stars with persistent disks remain slow rotators. Disk coupling as a way to moderate the angular momentum of the star as it settles onto the main sequence has previously been suggested to explain the bimodal rotational period distribution observed for solar-type stars in Galactic open clusters \citep{gallet2013,bouvier2014}. Observational tests suggested to narrow down the underlying scenario include measurements of the binary fractions, comparisons of the concentrations of slow and fast rotators, and \vsini{} measurements of stars with and without disks in star-forming regions. Spectroscopic binary studies have not revealed any significant differences between the binary fractions of slow and fast rotators thus far \citep{kamann2020,kamann2023,bu2024}, while the recent discovery of faster rotation among disk-less stars in the star-forming region NGC~2264 by \citet{bu2025} appears in line with the expectations of the disk-locking scenario proposed by \citet{bastian2020}. 

Another open question concerns the average equatorial velocities $\langle V_{\rm eq}\rangle$ of the cluster stars, relative to their expected critical (or break-up) velocities. Photometric studies trying to reproduce the CMDs of young or intermediate-age clusters with rotating stellar models commonly find that the models cannot satisfactorily reproduce the observed morphologies unless additional age spreads, high binary fractions, or rather extreme ($\gtrsim$95\% critical) rotation rates are assumed \citep[e.g.,][]{gossage2019,lipatov2022,cordoni2024}. Nearly critical rotation may also be favoured by the rich populations of Be stars, fast-rotating B stars with decretion disks that show Balmer-line emission \citep{rivinius2013}, which have been detected in young massive clusters via \ion{H}{$\alpha$} narrow-band photometry \citep{bastian2017,milone2018}. In the field, Be stars are commonly associated with binary interactions, following the spin up of a star via mass transfer from a companion \citep[e.g.,][and references therein]{bodensteiner2025}. However, \citet{hastings2020} showed that single stars born as fast rotators could also evolve into Be stars towards the end of their main sequence lifetimes, while the Be-star fractions observed among the eMSTO of some clusters of $\sim$50\% appear to outnumber the number of binary interaction products one can realistically expect in such stellar populations \citep{hastings2021}. The presence of large fractions of disk-embedded stars also has consequences on the photometric appearance of their host clusters, as absorption within the decretion disks may lead to stars appearing redder and fainter \citep[e.g.,][]{dantona2023,kamann2023}.

In light of the open questions concerning the origin of the distinctly rotating populations inside star clusters and their impact on our understanding of stellar evolution, additional observational constraints are urgently needed. Thanks to its spectroscopic multiplexing capabilities, MUSE offers the unique possibility to obtain \vsini{} measurements for large stellar samples and link them with high-quality \textit{Hubble} photometry. In this paper, we present MUSE observations of two massive LMC clusters, NGC~1866 ($\sim$200~Myr) and NGC~1856 ($\sim$300~Myr). Following a summary of our data reduction and analysis in Sec.~\ref{sec:data}, we survey the clusters for Be stars in Sec.~\ref{sec:emission} and present our stellar rotation measurements in Sec.~\ref{sec:rotation}. We discuss our results in the context of the origin of the stellar spin distribution and the cause of the Be phenomenon in Sec.~\ref{sec:discussion} before concluding in Sec.\ref{sec:conclusions}. Throughout the paper, we will discuss the clusters in order of their ages, rather than by their NGC numbers (i.e., NGC~1866 ahead of NGC~1856).

\section{Data}
\label{sec:data}

\subsection{Observations}
\label{subsec:observations}

We observed NGC~1866 and NGC~1856 with the Multi-Unit Spectroscopic Explorer \cite[MUSE,][]{muse} as part of the guaranteed time observations (GTO) programme ``A spectroscopic census of (Extra)galactic globular clusters with MUSE'' (PI: Dreizler, Kamann). We used the wide-field mode of MUSE, corresponding to a $1\times1$~arcmin field of view and a sampling of 0.2~arcsec. Ground-layer adaptive optics were used to enhance the image quality during the observations.

Observations of NGC~1866 were obtained during three nights (2019-11-30, 2019-12-26, 2020-02-22), while NGC~1856 was observed during four nights (2018-12-11, 2019-01-02, 2019-03-05, and 2021-10-03). Each observation consisted of $4\times600$~s exposures, with small spatial offsets and derotator offsets of 90~degrees in between them. All but the last observation of NGC~1856 were carried out using the nominal wavelength range of MUSE, covering 4~800 to 9~300~$\text{\AA}$. The remaining observation covered an extended wavelength range of 4~600 to 9~300~$\text{\AA}$.

\subsection{Reduction}
\label{subsec:reduction}

\begin{figure*}
 \includegraphics[width=\columnwidth]{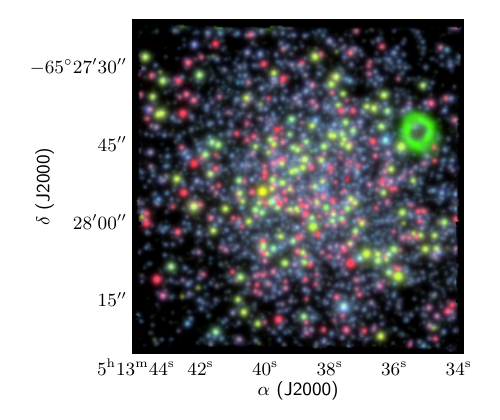}
 \includegraphics[width=\columnwidth]{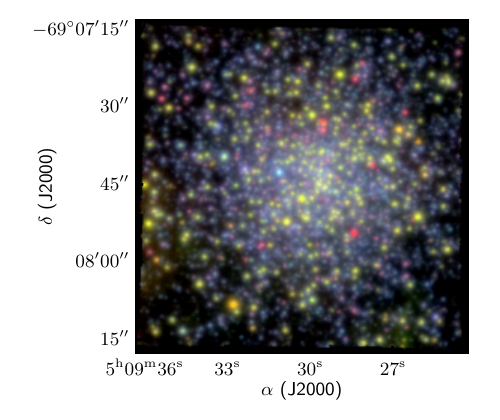}
 \caption{Colour-images of NGC~1866 (left) and NGC~1856 (right), created from the MUSE cubes with the best seeing available for each cluster. To highlight gas structures and emission-line stars, we extracted narrow-band images centred on \ion{H}{$\alpha$} (red), \ion{N}{ii}~6584 (green), and \ion{O}{iii} (blue) from each cube. In this representation, stars showing \ion{H}{$\alpha$} in emission appear red, whereas evolved giant stars appear green. Note the presence of a ring-shaped planetary nebula in NGC~1866, located in the north western quadrant of the MUSE field of view.}
 \label{fig:rgb}
\end{figure*}

We used the standard MUSE pipeline \citep{pipeline} to reduce the raw MUSE data. Each individual CCD image was bias-corrected before the signals of the individual spectra were traced, extracted, and wavelength calibrated. In addition, flat fielding and a correction for geometrical distortions were applied. The data from the 24 CCDs of each exposure were then combined and flux calibrated. Finally, the data of the four exposures performed per night and cluster were combined and resampled to a final data cube. For further details on the data reduction process, we refer to \citet{pipeline} and previous publications of our group \citep[e.g.][]{Kamann2018}.

In Fig.~\ref{fig:rgb}, we show colour images of both clusters, generated from the cubes that had the best seeing measured from the MUSE data (0.60~arcsec for NGC~1866, 0.64~arcsec for NGC~1856). To emphasize emission line sources as well as potential extended gas structures, the colour channels displayed in Fig.~\ref{fig:rgb} represent narrow-band images of 2~$\text{\AA}$ width around \ion{H}{$\alpha$}, \ion{N}{ii}~6584, and \ion{O}{iii}~5007, Doppler-shifted according to the systemic line-of-sight velocity of each cluster. Large numbers of stars with \ion{H}{$\alpha$} emission can be identified in both clusters according to their reddish colours. Furthermore, a bright ring-like structure is visible towards the north-western corner of the NGC~1866 field of view. This structure, which we classify as a planetary nebula, will be further discussed in a forthcoming publication (Bond et~al., in prep.). Finally, closer inspection of the NGC~1856 data revealed two low-intensity emission-line regions in NGC~1856, noticeable as areas with a greenish glow towards the south-west corner and the eastern edge in the right panel of Fig.~\ref{fig:rgb}.

\subsection{Photometry}
\label{subsec:photometry}

Our analysis is supported by two sets of \textit{Hubble} Space Telescope (HST) photometry. First, to perform the extraction of the stellar spectra from the MUSE cubes (cf. Sec.~\ref{subsec:analysis} below), we gathered HST/WFC3 images in F336W and F814W (proposal IDs: GO-13379, GO-14204) from the archive and processed them using \textsc{Dolphot} \citep{dolphot2000,dolphot2016}. We noticed that because of CCD saturation, some of the bright stars were missing in the reduced photometry. We compensated for this issue by cross-matching the HST photometry with \textit{Gaia} data release 3 and adding any Gaia sources to the catalogue that had no obvious counterpart.

However, most of the analysis presented in the following uses the astrometric catalogues made available by \citet{Niederhofer2024}. For both clusters, \citet{Niederhofer2024} provide photometry in four filters, namely F336W, F438W, F555W and F814W, using additional data from the GO-13011, GO-14069 and GO-16748 programmes. The data have been corrected for differential reddening, applying the techniques described by \citet{Milone23}. Briefly, we first determined for each star the extinction coefficient $A_{\lambda}$ in each filter. Thereby, we took into account the dependence of $A_{\lambda}$ on the spectral type of the stars. To this end, we created for each filter a grid of extinction coefficients across a broad range of stellar temperatures by comparing synthetic stellar spectra with and without reddening, convolved with the respective filter transmission curves. We assumed a \citet*{Cardelli89} reddening law with R$_V$=3.1. For each star in the catalogue, we then determined an estimate of its effective temperature as the temperature of the closest model point (in colour-magnitude space) of a BaSTI isochrone \citep{hidalgo2018} that resembles best the cluster sequences. Then, we selected for each cluster a sample of well-measured main-sequence stars as reference objects for the determination of the local reddening. We chose stars well below the eMSTO feature and excluded stars on the binary sequence. The adopted technique to map the reddening across the cluster relies on taking into account simultaneously the information coming from all filters. Thus, we constructed for each cluster three different CMDs: $m_{\lambda}-m_{\rm F814W}$ vs $m_{\rm F814W}$, where $\lambda$ corresponds to F336W, F438W and F555W, respectively. Within each CMD, we derived the mean ridge-line of the reference stars, by first dividing the magnitude range covered by the reference stars into bins of 0.2~mag and then computing the sigma-clipped median colour within each bin. These median colours are then fit with a cubic spline model.  
Subsequently, we determined for each reference star its distance from the ridge line in the direction of the reddening vector. We then compared these distances coming from all CMDs with predictions derived for a range of E(B$-$V) values, using the corresponding extinction coefficients. Specifically, we created an array of E(B$-$V) values, ranging from $-$0.3~mag to 0.3~mag in steps of 0.001~mag and chose as the best-fitting reddening the one that provides the minimum $\chi^2$. 
Finally, we determined the local reddening for each star in the catalogue as the 2.5$\sigma$-clipped median value of the nearest 75 reference stars, excluding the target star itself. 
While differential reddening is negligible for NGC~1866, we found significant differences across the observed field in NGC~1856, with a peak-to-peak variation of the reddening of $\delta$E(B$-$V)$\sim$0.2~mag. 

We used the proper motions measured by \citet{Niederhofer2024} to select likely members of NGC~1866 and NGC~1856. To this aim we determined the expected dispersion of cluster stars as a function of F814W magnitude. The dispersion consists of the intrinsic velocity dispersion of the clusters stars $\sigma_{\rm int}$, and the dispersion due to the associated errors of the proper motion measurements $\sigma_{\rm err}$. For $\sigma_{\rm int}$, we assumed 2.64~km\,s$^{-1}$ and 2.31~km\,s$^{-1}$ for NGC~1856 and NGC~1866, respectively \citep{McLaughlin05}. At the distance of the LMC, these values translate to $\sim$0.01~mas\,yr$^{-1}$. We then determined as $\sigma_{\rm err}$ the median proper motion measurement error as a function of the F814W magnitude. The total dispersion is then calculated as $\sigma_{\rm tot} = \sqrt{\sigma_{\rm int}^2 + \sigma_{\rm err}^2}$. We kept as likely cluster members those stars that are within 2.5 times the total dispersion at the respective magnitudes of the stars. 

In Fig.~\ref{fig:cmd}, we show the reddening-corrected and proper-motion-cleaned ($m_{\rm F336W}-m_{\rm F438W}$ vs $m_{\rm F438W}$) CMDs of the two clusters. To better visualise the colour spreads along the main sequences of NGC~1866 and NGC~1856, we determined their blue and red envelopes as follows. First, we discarded evolved stars by imposing a colour cut of $m_{\rm F336W} - m_{\rm F438W}<0.4$. The remaining sample was split up into half-overlapping $m_{\rm F438W}$ bins of $0.5~\text{mag}$ width. Within each bin, we determined the median ($m_{\rm F336W}-m_{\rm F438W}$) colour, and cleaned the sample from outliers by removing all stars with colour offsets $>0.3$ from the median, followed by kappa-sigma clipping. Then, we measured the 5th and 85th percentiles of the $m_{\rm F336W} - m_{\rm F438W}$ colour distribution. The latter percentile was chosen in order to approximately exclude the binary sequence located towards the red of the (single-star) main sequence. Finally, the blue and red envelopes were determined by interpolating between the percentile values in each magnitude bin. The lines determined this way are included in the left panels of Fig.~\ref{fig:cmd}. They were used to verticalise the main sequences of both clusters, as illustrated in the right panels of Fig.~\ref{fig:cmd}. In both clusters, we recover the split main sequence, most prominently in the magnitude range $19<m_{\rm F438W}<21$.

\subsection{Spectrum extraction and combination}
\label{subsec:analysis}

\begin{figure}
    \centering
    \includegraphics{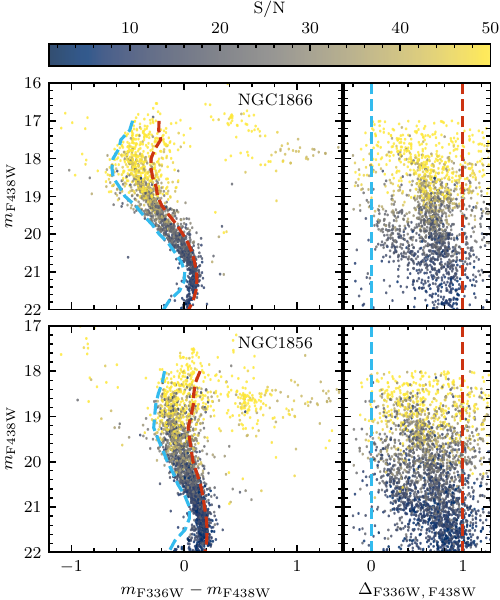}
    \caption{HST colour-magnitude diagrams of NGC~1866 (top) and NGC~1856 (bottom). The left panels show the full cluster populations, whereas the right panels zoom into the main sequence regions. Dashed blue and red lines indicate the fiducial lines that have been used to verticalize the main sequence of either cluster (see text for details). In all panels, we only show the stars with available spectroscopy, colour-coded according to their S/N.}
    \label{fig:cmd}
\end{figure}

We processed the seven data cubes with \textsc{PampelMuse} \citep{pampelmuse}. Based on an astrometric reference catalogue, \textsc{PampelMuse} determines the point spread function (PSF) and the positions of the stars that are bright enough to be resolved as a function of wavelength. This information is then used to deblend and extract individual object spectra from the cubes. The astrometric reference catalogues required for this task were generated from archival HST data as outlined in Sec.~\ref{subsec:photometry}. 

In order to maximise the signal-to-noise ratio (S/N) of the individual stars' spectra, we combined the spectra extracted from the single-epoch MUSE cubes on a star-by-star basis. In the combination process, we discarded spectra with a magnitude accuracy (\texttt{MAG ACCURACY} parameter determined by \textsc{PampelMuse}) of $<0.5$ and a S/N of less than 50\% of the highest value obtained for each star. Furthermore, we omitted all spectra with an active \texttt{MULTISTAR} flag, indicating that the extracted spectrum contained flux contributions from more than one star. During the combination, the spectra were weighted by their S/N. Given the constant wavelength sampling of the MUSE data, no resampling was needed.

We note that we did not account for any binary-induced line-of-sight velocity variations in between the different epochs when combining the spectra. Upon analysis of the single-epoch spectra, we found that for both clusters, the fraction of stars showing peak-to-peak velocity variations that are large enough to inflate our \vsini{} measurements ($>$50~\kms{}) is below 10\%.

Uncertainties for the combined spectra were obtained via standard error propagation of the uncertainties determined by \textsc{PampelMuse}. The latter tend to underestimate the true uncertainties because the resampling of the MUSE data during the reduction creates covariances between voxels that are not propagated by the MUSE pipeline \citep[see Sec. 4.6 of][]{pipeline}. Therefore, we applied an a posteriori correction to the uncertainties calculated for the combined spectra. For the stars with at least three valid input spectra, we determined the standard deviation of the continuum-subtracted input spectra as a function of wavelength and divided the result by the formally propagated uncertainties. Once all spectra were processed, we determined the median of these correction factors and multiplied it by the formally propagated uncertainties.

We obtained combined spectra for 2\,531 stars in NGC~1866 and 4\,406 stars in NGC~1856. Their distribution across the CMDs of both clusters is visualised in Fig.~\ref{fig:cmd}, colour-coded by the S/N of the stars in the MUSE data. Using isochrones (c.f., Sec.~\ref{subsec:spexxy}), we estimate that the stellar masses covered by our samples range from 1.5~\msun{} to 3.1~\msun{} (NGC~1856) and 3.7~\msun{} (NGC~1866).

\subsection{Full-spectrum fitting}
\label{subsec:spexxy}

The combined spectra were analysed with \textsc{Spexxy} \citep{spexxy}, which performs a full spectrum fitting against a library of template spectra in order to infer stellar parameters, line-of-sight velocities, and line broadening. As templates, we adopted the synthetic spectra from \citet{ferre}. The template spectra were convolved with a model for the wavelength-dependent MUSE line-spread function (LSF) prior to the analyses.

Initial estimates of surface gravity ($\log g$) and effective temperature ($T_{\rm eff}$) were derived by comparing the HST photometry of either cluster to isochrones from the \texttt{MIST} data base \citep{mist_i,mist_ii}. When generating the isochrones, we assumed $Z=0.006$ and $m-M=18.3$ for both clusters, and extinction values of $A_{\rm V}= 0.37$ (NGC~1866) and $0.59$ (NGC~1856), based on \citet{milone2018}. The isochrones were generated for a range of ages, from $\log ({\rm age}/{\rm yr})=8.0$ to 8.5, using a stepsize of 0.05. We visually determined the isochrone yielding the best fit to the photometry of each cluster, finding $\log ({\rm age}/{\rm yr})=8.3$ for NGC~1866 and 8.5 for NGC~1856. 

As initial guesses for the metallicity $[{\rm M/H}]$ and the line-of-sight velocity $v_{\rm LOS}$, we adopted constant values of -0.5 and $260~{\rm km\,s^{-1}}$, respectively, roughly matching the values measured by \citet{usher2019} from integrated-light spectroscopy. Finally, spectral fits used to constrain stellar rotation velocities were initialised with a value of $V\sin i=100\,{\rm km\,s^{-1}}$. We note that our approach to constrain $V\sin i$ differs from our previous works \citep{kamann2020,kamann2023}, where we used a Gaussian line-of-sight velocity distribution to account for any line broadening and applied an a posteriori correction to convert the fitted dispersion values to $V\sin i$. Instead, for the current analysis, we modified \textsc{Spexxy} such that $V\sin i$ could be fitted directly. To this aim, we implemented a recipe for rotational line broadening in \textsc{Spexxy}, using the functions provided by the \textsc{PyAstronomy}\footnote{\url{https://github.com/sczesla/PyAstronomy}} \citep{pyastronomy} package which are based on the formulae of \citet{gray2005}.

In general, we tried to constrain the following parameters through the spectral fits with \textsc{Spexxy}: $\log g$, $T_{\rm eff}$, $[{\rm M/H}]$, $v_{\rm LOS}$, and \vsini{}. However, depending on the expected properties of the star at hand, we decided to fix certain parameters during individual fits. First, given the low spectral resolution, $\log g$ was always fixed when an initial value from the isochrone was available. In such cases, we used a threshold in the initial surface gravity of $\log g=3.2$ to separate main sequence from evolved stars. This division was motivated by the fact that evolved stars have rotational velocities below the sensitivity of MUSE \citep[see][]{kamann2020}, whereas the hot main-sequence stars in YMCs lack the metallic spectral lines required for spectroscopic $[{\rm M/H}]$ determinations. Consequently, we fixed the rotational broadening to zero for the former and $[{\rm M/H}]$ to the initial guess for the latter.
Finally, stars with signs of \ion{H}{$\rm \alpha$} emission (see next section) were treated like main-sequence stars, but the fit was limited to the spectral region around the Paschen lines ($\geq8\,400\,{\text \AA}$). However, because the Paschen series can also be impacted by line emission, we discuss the reliability of these fits in Sec.~\ref{sec:stellar_rotation_across_the_CMD} below.

\subsection{Spectroscopic sample selection}
For the following analyses, we selected all combined spectra of cluster members (cf. Sec.~\ref{subsec:photometry}) with a magnitude accuracy $>0.6$, $\mathrm{S/N}>20$, and a formally successful (i.e., converged) \textsc{Spexxy} fit. These quality criteria were passed by 1\,212 stars in NGC~1866 and 1\,357 stars in NGC~1856. To further restrict the samples to main-sequence stars when necessary, we used the verticalized CMDs displayed in the right panels of Fig.~\ref{fig:cmd} and required pseudo-colours in the range $-0.2<\Delta_{\rm F336W,\,F438W} < 1.2$.

\section{Emission-line stars}
\label{sec:emission}

There is substantial value in the detection and classification of stars with emission lines in young star clusters, because line emission is commonly associated with stars on non-standard evolutionary pathways, such as interacting binaries and critically rotating (Oe or Be) stars. Thanks to its unique combination of line sensitivity and multiplexing capabilities, MUSE has proved to be a powerful tool for characterising emission-line stars in young Magellanic Cloud clusters \citep[e.g.,][]{bodensteiner2020,bodensteiner2021,saracino2023}.

 The analysis of \citet{dupree2017} already showed that NGC~1866 hosts a number of Be stars among its main sequence population. Similarly, \citet{kamann2023} spectroscopically uncovered a large population of such stars in NGC~1850. Photometric studies of massive and open clusters with ages less than a few hundred Myr have also detected large populations of Be stars within them \citep[e.g.,][]{bastian2017, milone2018}, highlighting their ubiquity in young star clusters.

\subsection{Identification}
\label{subsec:emission_identification}

\begin{figure}
    \centering
    \includegraphics{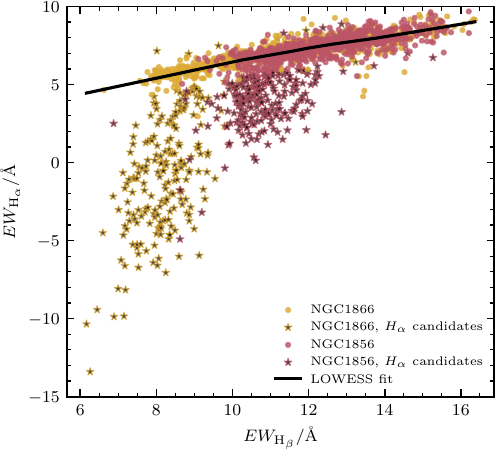}
    \caption{Relation between the equivalent widths of \ion{H}{$\alpha$} and \ion{H}{$\beta$} for the main sequence samples in NGC~1866 (orange) and NGC~1856 (red). The black solid lines shows the LOWESS fit to the data with $3<EW_{\rm H_{\alpha}}/\text{\AA}<9$. Outliers as defined in the text are shown as open stars.}
    \label{fig:equivalent_width}
\end{figure}

\begin{figure*}
    \includegraphics{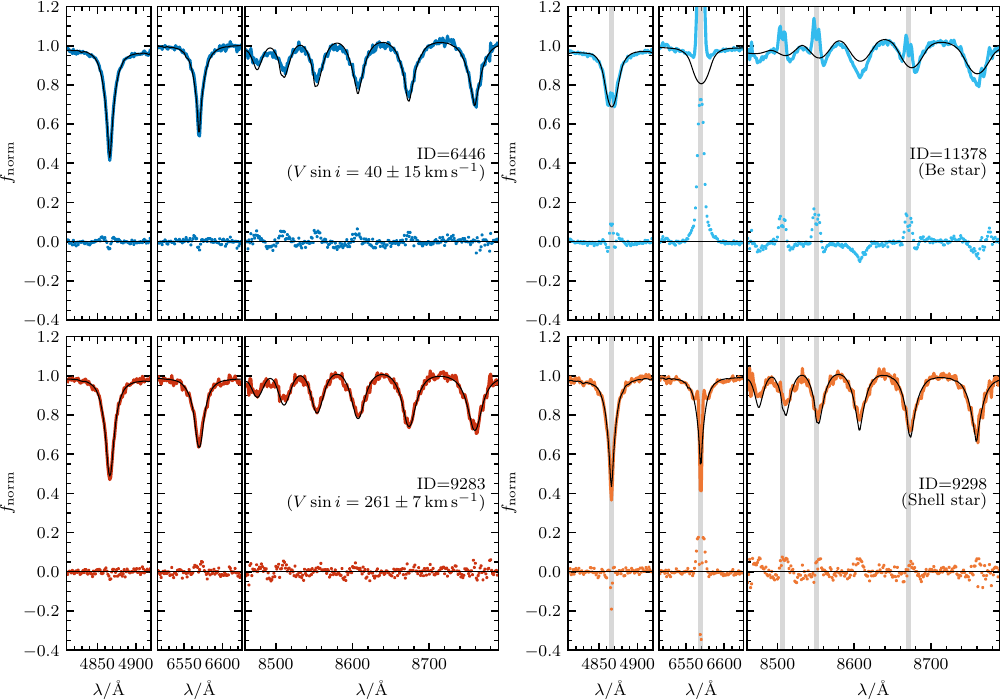}
    \caption{Normalized example spectra from the NGC~1866 sample around \ion{H}{$\beta$}, \ion{H}{$\alpha$}, and the Paschen series. The left panels show a slow (top) and a fast (bottom) rotator without emission lines. Our \vsini{} measurements are provided together with the star IDs. In the right panels, two types of emission-line star are presented, a Be star at the top and a shell star at the bottom. The locations of different emission features are marked in grey. Note that, besides Balmer emission, the Be star also shows prominent, double peaked \ion{Ca}{ii} triplet emission. In all panels, the observed spectrum is shown as a thick coloured line, while the best fit is indicated by a narrow black line. The fit residuals are provided towards the bottom of each panel.}
    \label{fig:be_spectra}
\end{figure*}

To homogeneously detect emission line stars within our spectroscopic samples, we followed a two-step process.
First, we measured the equivalent widths (EWs) of \ion{H}{$\alpha$} and \ion{H}{$\beta$} in the combined spectra and looked at their relation, shown in Fig.~\ref{fig:equivalent_width}. While most stars follow a linear relation as expected, a significant fraction in either cluster shows considerably weaker \ion{H}{$\alpha$} lines, or net \ion{H}{$\alpha$} emission (i.e., negative $EW_{\rm H_{\alpha}}$). To identify potential emission-line stars, we performed locally-weighted scatterplot smoothing \citep[LOWESS,][]{lowess} using the \texttt{python} package \texttt{statsmodels} \citep{statsmodels} and considered as candidates all stars for which the equivalent width of \ion{H}{$\alpha$} deviated by $>1\,{\text\AA}$ and $>5\times$ its uncertainty. The 448 candidates are shown as open stars in Fig.~\ref{fig:equivalent_width}.\footnote{We note that seven sources are identified as outliers towards higher EW in Fig.~\ref{fig:equivalent_width}. Relative to the 441 outliers towards lower EW, they indicate that the number of statistical interlopers is $\sim$1-2\%.}

During the second step, we utilized the residuals of the \textsc{Spexxy} full-spectrum fits around the \ion{H}{$\alpha$} lines to search for evidence of single- or double-lined emission components (see Fig.~\ref{fig:be_spectra}). To this aim, we fitted the \ion{H}{$\alpha$} residuals as a function of wavelength using three models, (1) a linear fit, (2) a single Gaussian, and (3) a double Gaussian, with both components sharing the same centroid but differing in the signs on their amplitudes. The inclusion of a double Gaussian model is motivated by our previous study of the 100~Myr old cluster NGC~1850 \citep{kamann2023}, which hosted many shell stars, Be stars observed through their decretion disks (see Sec.~\ref{sec:shell_stars}) and characterized by double-peaked \ion{H}{$\alpha$} emission. All fits were performed using \textsc{lmfit} \citep{lmfit}. The best-fitting model was selected using the Akaike inference criteria (AIC). In the case of the single Gaussian, we further required that its centroid was within $1~\text{\AA}$ of the line centre expected according to the line-of-sight velocity measured by \textsc{Spexxy} and that its amplitude was detected at $3~\sigma$ significance. The same wavelength criterion was imposed for the (combined) centroid of the double Gaussian. In addition, each of its components needed to be detected at $3~\sigma$ significance and contribute at least 20\% to the total flux of the model. The last criterion was adopted to avoid that deviations from a Gaussian line shape result in spurious shell-star detections, following visual inspection of some spectra.

\begin{figure}
    \centering
    \includegraphics{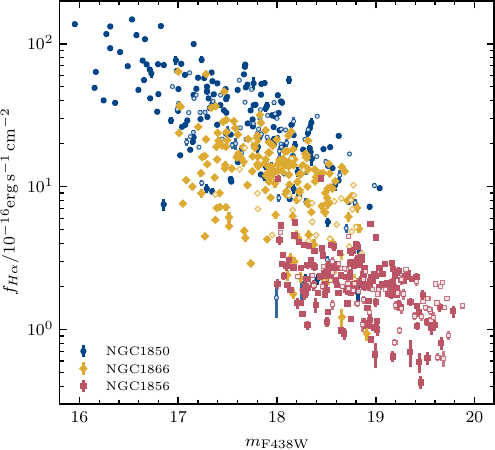}
    \caption{\ion{H}{$\alpha$} emission-line fluxes of Be stars in the MUSE samples of NGC~1850 \citep[blue circles,][]{kamann2023}, NGC~1866 (orange diamonds) and NGC~1856 (red squares) stars as a function of $m_{\rm F438W}$ magnitude. The \ion{H}{$\alpha$} fluxes have been measured by fitting Gaussian profiles to the \textsc{spexxy} fit residuals as described in the text. Double-peaked \ion{H}{$\alpha$} lines which were best fitted by double-Gaussian profiles are shown as open symbols.}
    \label{fig:halpha_fluxes}
\end{figure}

\subsection{Demographics}
\label{subsec:emission_demographics}

Stars for which the spectra were identified as outliers in the relation shown in Fig.~\ref{fig:equivalent_width} and the fit residuals were best fit by either one of the Gaussian models were considered as Be stars. Example spectra for two such stars (one showing single-peak, one double-peak \ion{H}{$\alpha$} emission) from the NGC~1866 sample are available in Fig.~\ref{fig:be_spectra}. Using the amplitudes of the Gaussian fits, we show their \ion{H}{$\alpha$} emission fluxes as a function of magnitude in Fig.~\ref{fig:halpha_fluxes}. In cases where a double Gaussian profile yielded the best fit to the \ion{H}{$\alpha$} residuals, we considered the amplitude of the emission component as representative for the true (i.e., unabsorbed by the decretion disk) flux. Such spectra are represented by open symbols in Fig.~\ref{fig:halpha_fluxes}. To increase the stellar mass range over which the \ion{H}{$\alpha$} fluxes can be studied, we performed the same analysis on the sample of Be stars identified by \citet{kamann2023} in NGC~1850 and included the corresponding line fluxes in Fig.~\ref{fig:halpha_fluxes}.

A few aspects can be highlighted in Fig.~\ref{fig:halpha_fluxes}. First, a clear correlation exists between the magnitude of a star and its \ion{H}{$\alpha$} flux, with brighter stars showing higher emission line fluxes. Interestingly, this relation extends over all three clusters, with only a small shift towards lower fluxes visible for the NGC~1856 stars at fixed $m_{\rm F438W}$ magnitude.

When comparing single- and double-peaked line emitters within each cluster, there is a trend for the latter to have fainter $m_{\rm F438W}$ magnitudes on average. As will be further discussed in Sec.~\ref{sec:rotation}, we also find that these shell-star candidates show redder colours than the bulk of the main sequence stars. Self-extinction by the decretion disks appears as a probable explanation for these two effects. We also note that both NGC~1866 and NGC~1856 display some bimodality, in the sense that at fixed $m_{\rm F438W}$, there seem to be two sequences separated by a factor $\sim$2-3 in \ion{H}{$\alpha$} flux. Most stars follow the upper sequence. The origin of this bimodality is currently unclear.

\begin{figure}
    \centering
    \includegraphics{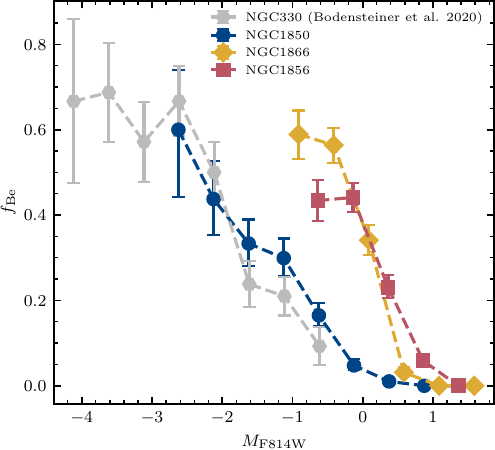}
    \caption{Fraction of Be stars (relative to the total number of main sequence stars) as a function of absolute F814W magnitude in young massive Magellanic Clouds clusters: NGC~330 \citep[grey hexagons, from][]{bodensteiner2020}, NGC~1850 (blue circles), NGC~1866 (orange diamonds), and NGC~1856 (red squares). Uncertainties for the latter three clusters have been estimated from the sample sizes per magnitude bin, using the \citet{Wilson1927} score interval.}
    \label{fig:be_fraction}
\end{figure}

Fig.~\ref{fig:halpha_fluxes} shows that the population of line emitters in NGC~1856 extends to fainter magnitudes, despite the similar characteristics relative to the other two clusters in terms of mass, metallicity, or main-sequence morphologies. The observation that in the older cluster, lower-mass stars show \ion{H}{$\alpha$} emission suggests that the stars ``switch on'' their \ion{H}{$\alpha$} emission towards the end of their main-sequence lifetimes. This scenario is also supported by Fig.~\ref{fig:be_fraction}, which shows the fraction of Be stars ($f_{\rm Be}$) as a function of magnitude for both clusters studied in this work, NGC~330 \citep[30~Myr, see][]{bodensteiner2020}, and NGC~1850 (100~Myr). For the latter cluster, we recalculated the $f_{\rm Be}$ trend reported in \citet{kamann2023} as a function of $m_{\rm F814W}$. Further, to ease the comparison between the four clusters, we converted $m_{\rm F814W}$ to absolute F814W magnitudes, using the distance moduli and reddening values from \citet{Milone23}, and an extinction coefficient of $A_{\rm F814W}=1.88\,E_{(B-V)}$. We see that the occurrence of emission-line stars is largely confined to the brightest two magnitudes in all clusters and increases sharply towards the main sequence turn-off. Similar trends have also been observed in photometric studies of YMCs in the Magellanic Clouds \citep[e.g.,][]{bastian2017,milone2018}, where the identification of Be stars was based on colour excesses in \ion{H}{$\alpha$} narrow-band filters.

In all four clusters studied spectroscopically, which have turn-off masses between 3.1~\msun{} (NGC~1856) and 7.5~\msun{} (NGC~330), we observe high emission line fractions of $\sim$50-60\% around the turn-off. For NGC~330, NGC~1850 and NGC~1856, we can compare these fractions to the photometric results of \citet{milone2018}. While for NGC~330 and NGC~1850 there is an overall agreement between the two studies, \citet{milone2018} find a significantly lower fraction of emission-line stars of $<20$\% in NGC~1856. We argue that the discrepancy can be attributed to the higher sensitivity of spectroscopic searches for line emission compared to narrow-band photometry. Indeed, Fig.~\ref{fig:halpha_fluxes} demonstrates that the typical \ion{H}{$\alpha$} fluxes in NGC~1856 are considerably lower than in the younger clusters. 
\citet{dupree2017} noted another problem with photometric Be star searches, namely that the \ion{H}{$\alpha$} line will gradually shift out of the transmission window of the narrow-band filter with increasing line-of-sight velocity. NGC~1856 has a systemic velocity around 300~\kms, and our Gaussian fits yield median line widths of $\sigma\sim3.3\,\text{\AA}$. Comparing to the transmission curve of the WFC3-F656N filter onboard HST, which displays a sharp cut-off beyond $6570\,\text{\AA}$, we estimate that about 30\% of the line flux is missed by the narrow-band filter.
All in all, our results suggest that photometric searches for emission line stars may significantly underestimate their fractions in clusters older than $\sim100~{\rm Myr}$.

\subsection{Shell stars}
\label{sec:shell_stars}

As already touched upon above in Sec.~\ref{subsec:emission_demographics} and illustrated in Fig.~\ref{fig:be_spectra}, Be stars showing sharp absorption cores in their \ion{H}{$\alpha$} profiles are synonymous with shell stars, Be stars that are observed at high inclination angles ($\sin i\sim1$, i.e. equator-on), such that our view onto the star is (partially) blocked by its decretion disk. This self-extinction does not only affect the \ion{H}{$\alpha$} profile of the star, but can also affect its photometry, such that it appears redder and fainter \citep{rivinius2013}. Furthermore, the spectra of shell stars frequently show other peculiarities, such as \ion{Fe}{ii} and \ion{Si}{ii} absorption or narrow high-order Paschen lines. However, all these features vary substantially in strength, and their presence in the MUSE spectra is often not obvious, complicating a consistent distinction between shell stars and classical Be stars.

The demographics of the \ion{H}{$\alpha$} line contain important information about the morphologies of the decretion disks. In particular, they allow us to estimate the disk half-opening angle, i.e., the average angle between the equatorial plane and the surface of the disk as measured from the star. Using the criteria of Sec.~\ref{subsec:emission_identification}, we find that the \ion{H}{$\alpha$} lines of 34/208 Be stars in NGC~1866 and 59/221 Be stars in NGC~1856 are best fit by double-peaked profiles, corresponding to shell-star fractions of $16^{+6}_{-5}$\% and $26^{+6}_{-6}$\%, respectively. The values agree well with the fractions of 23\% determined for NGC~1850 \citep{kamann2023} and 22.8\% determined for the galactic field \citep{hanuschik1996}. Under the assumption of isotropic spin axis distributions, the disk half-opening angles are $9.2^{+3.0}_{-3.5}$~degrees for NGC~1866 and $15.1^{+3.6}_{-3.6}$~degrees for NGC~1856. One may speculate about an age or mass evolution in the sense that older, lower-mass Be stars have fluffier disks. However, more data are needed to investigate this, also considering that the opening angle for NGC~1850 was estimated to be 13~degrees, despite the cluster being younger than both NGC~1866 and NGC~1856.

\section{Stellar rotation}
\label{sec:rotation}

\subsection{Stellar rotation across the CMD}
\label{sec:stellar_rotation_across_the_CMD}

\begin{figure*}
    \centering
    \includegraphics{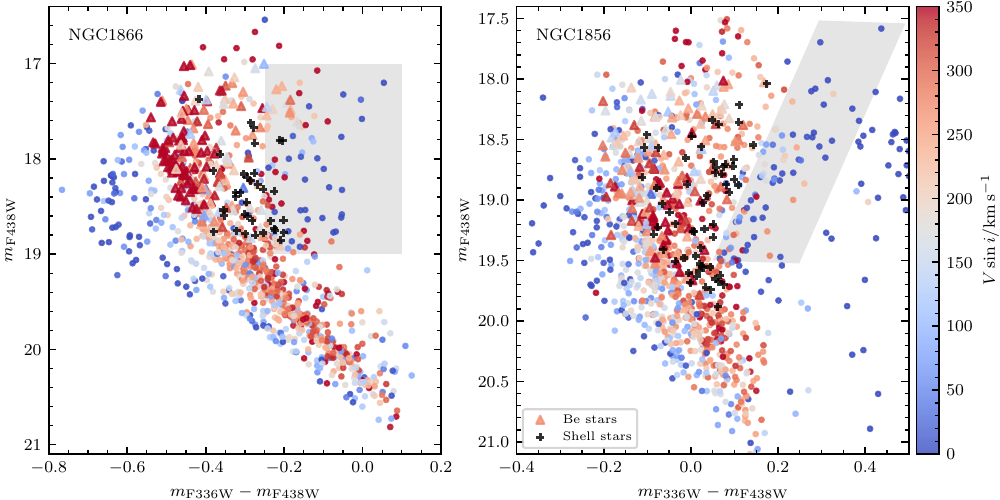}
    \caption{Distribution of $V\sin i$ along the main sequences of NGC~1866 (\textit{left}) and NGC~1856 (\textit{right}). In both panels, stars are colour-coded according to their measured $V\sin i$. Be stars are highlighted using triangles, while shell stars, for which no $V\sin i$ values are reported, are plotted using black crosses. The slowly rotating stars forming a red extension to the eMSTO of either cluster are highlighted using grey polygons (see text for details).}
    \label{fig:cmd_vsini}
\end{figure*}

In Fig.~\ref{fig:cmd_vsini}, we show colour-magnitude diagrams of NGC~1866 and NGC~1856, focussing on the main sequences of the two clusters and colour-coding the data points by the $V\sin i$ values measured for the corresponding stars. In both cases, we observe a correlation between the colour of a star and its projected rotation velocity, in the sense that redder stars rotate faster (at fixed $m_{\rm F438W}$). This behaviour is in agreement with what has been observed in other young clusters of the LMC, such as NGC~1850 \citep{kamann2023} and NGC~1818 \citep{marino2018}. Stars on the blue main sequence are characterised by $V\sin i$ values $<100~{\rm km\,s^{-1}}$, whereas the red main-sequence stars appear to cluster around $200~{\rm km\,s^{-1}}$.

As expected, the Be stars identified in Sec.~\ref{sec:emission} are fast rotators according to our fits. It is interesting to note that their $V\sin i$ values appear to exceed those of the red main-sequence stars. We will come back to this point below. Also shown in Fig.~\ref{fig:cmd_vsini} are the shell stars we identified based on the double-peaked \ion{H}{$\alpha$} emission lines visible in the MUSE spectra. We do not report their $V\sin i$ values in Fig.~\ref{fig:cmd_vsini} because their spectra are often dominated by narrow absorption lines originating from self-absorption in their decretion disks, which are not reproduced by our template spectra.

We further notice in Fig.~\ref{fig:cmd_vsini} a population of predominantly slow rotators that form an extension of the eMSTO of each cluster toward redder colours. The approximate locations of these stars are indicated by the grey polygons plotted in each panel of Fig.~\ref{fig:cmd_vsini}. It is not immediately obvious what these stars could be. In stellar evolutionary models that aim to reproduce the eMSTOs of young clusters \citep{georgy2013,gossage2019}, there is no equivalent population of red, yet slowly rotating stars (although isochrones initiated with different rotation speeds do cross near the turn-off). In the analysis of NGC~1850 stars \citep{kamann2023}, we found that shell stars predominantly populated the red extension of the eMSTO. While these stars are fast rotating, the characteristics of their spectra lead to spuriously low $V\sin i$ values during the analysis of the MUSE spectra (see above). However, we consider it unlikely that the slow rotators seen towards the red of the eMSTOs of NGC~1866 and NGC~1856 are shell stars, because none of them show \ion{H}{$\alpha$} emission. We also discard the possibility that these stars are fast rotators observed pole-on, because fast rotating stars show a positive effective temperature gradient from the equator to the poles \citep[e.g.,][]{vonzeipel1924}, such that they should appear bluer when observed at low inclination ($\sin i\sim0$). Interestingly, the photometric locations of these stars resemble those of UV-dim stars identified in other clusters \citep{martocchia2023,milone2023}. While their origin is still unclear, a recent study by \citet{leanza2025} shows that they are slowly rotating in the intermediate-age cluster NGC~1783.

\begin{figure}
    \centering
    \includegraphics{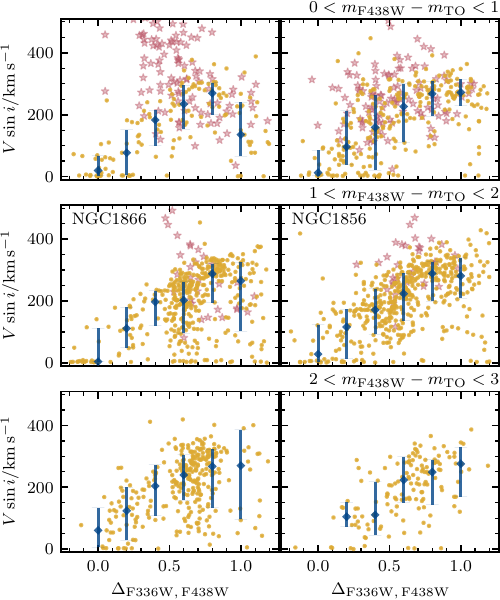}
    \caption{$V\sin i$ measurements as a function of colour, for different magnitude bins in NGC~1866 (\textit{left}) and NGC~1856 (\textit{right}). Magnitude bins are defined relative to the MSTO of each cluster, as indicated to the top right of each row. Orange circles show the individual measurements, while the median values together with the 16th and 84th percentiles within colour bins of $0.2~{\rm mag}$ are shown as blue diamonds. Be stars, shown as red stars, were not included when calculating the bin averages.}
    \label{fig:vsini_vs_colour}
\end{figure}

To study the rotational properties of the stars along the main sequences of the two clusters in more detail, we show in Fig.~\ref{fig:vsini_vs_colour} the individual $V\sin i$ measurements as a function of colour for different bins in $m_{\rm F438W}$ magnitude. To account for the age difference between the two clusters, we defined the magnitude bins relative to the MSTO of each cluster. As turn-off magnitudes, $m_{\rm TO}$, we selected the brightest F438W data points from the MIST isochrones (cf. Sec.~\ref{subsec:spexxy} that had a \texttt{phase} flag of 0 (indicating main sequence stars), resulting in $m_{\rm TO,\,NGC~1866}=17.37$ and $m_{\rm TO,\,NGC~1856}=18.17$, respectively. Further, to facilitate the comparison between the different bins and the two clusters, we determined the pseudo-colour $\Delta_{\rm F336W,\,F438W}$ for each star, defined as the $(m_{\rm F336W}-m_{\rm F438W})$ colour distance between each star and the blue ridge line defined in Sec.~\ref{subsec:photometry} (cf. Fig.~\ref{fig:cmd}), normalised by the colour difference between the red and blue ridge lines.

A general trend is apparent in Fig.~\ref{fig:vsini_vs_colour}, in that the mean $V\sin i$ increases with $\Delta_{\rm F336W,\,F438W}$, as expected. However, several aspects of the distributions shown in Fig.~\ref{fig:vsini_vs_colour} deserve further investigation. First, we note that the relation between the projected rotation velocity and photometric colour is remarkably similar across the different magnitude bins investigated and also across the two clusters in our sample. The bluest stars are generally slow rotators, with $V\sin i\lesssim 50\,{\rm km\,s^{-1}}$. Going towards redder colours, $V\sin i$ gradually increases up to typical values of $\sim250-300\,{\rm km\,s^{-1}}$ for stars with $\Delta_{\rm F336W,\,F438W}\gtrsim0.6$. This coherent picture suggests that the rotational properties of the stars remain largely unchanged across the mass and age ranges studied here. We remind the reader that our sample includes stellar masses between $\sim2.2\,{\rm M_\odot}$ in the faintest magnitude bin ($20<m_{\rm F438W}<21$) and $\sim3.7\,{\rm M\odot}$ at the eMSTO of NGC~1866.

The gradual increase of $V\sin i$ visible in Fig.~\ref{fig:vsini_vs_colour} appears at odds with the expectation from a bimodal distribution of rotational velocities\footnote{Which in turn may be expected due to the bi-modal colour distribution across the main-sequence.} in the two clusters, in which case one might expect to see a sharp increase in the average $V\sin i$ when moving from the blue to the red main sequence. A possible explanation for this trend could be a dependence of the observed colour on the inclination of a star, because for increasingly fast rotating stars, fundamental stellar parameters like effective temperature and surface gravity are expected to depend on latitude, in the sense that the equator region of the star appears cooler and fainter than the poles, an effect known as gravity darkening \citep{vonzeipel1924}. However, for clusters with ages similar to those studied in this work, \citet{gossage2018} studied the effect of gravity darkening using MESA models and found it to be strongest among turn-off stars, whereas colour variations among the main sequence stars were restricted to $<0.05\,{\rm mag}$ in B-V \citep[see also][]{wang2023}. Our measurements also suggest that gravity darkening only plays a small role in defining the colours of the main-sequence stars in NGC~1866 and NGC~1856. To illustrate this, consider the magnitude range $1$ to $2$~mag below the turn-off in NGC~1866. In this case, the red main-sequence stars occupy the pseudo-colour range $0.6<\Delta_{\rm F336W,\,F438W}<1$ (cf. Fig.~\ref{fig:cmd}). The central left panel of Fig.~\ref{fig:vsini_vs_colour} shows that the \vsini{} distribution of these stars extends down to zero at virtually no colour change. Hence, our results rather indicate that the assumption of a purely bimodal distribution of rotational velocities in young massive clusters is a simplification, and that their true stellar spin distribution is more complex.

In Fig.~\ref{fig:vsini_vs_colour}, the samples of Be stars identified in Sec.~\ref{sec:emission} are highlighted using red stars. We note that their $V\sin i$ values extend to higher values compared to ordinary main sequence stars, particularly in the case of NGC~1866. Although Be stars are expected to be fast rotating, in some cases, our measured values exceed the critical velocities typically assumed for such stars. For example, the SYCLIST models \citep{georgy2013} predict critical velocities around $400\,{\rm km\,s^{-1}}$ for main sequence stars with masses $\sim3\,{\rm M_\odot}$ after 200-300~Myr of evolution (see Sec.~\ref{subsec:red_main_sequence}). To understand these results, we visually inspected the spectral fits of the Be stars individually. We found that many spectra displayed features that have potentially biased our $V\sin i$ measurements, such as faint emission components in the Paschen lines or \ion{Ca}{ii} triplet emission. An example is shown in Fig.~\ref{fig:be_spectra}. Be stars are known to show emission lines apart from the Balmer lines \citep[e.g.,][]{banerjee2021}, yet the origin of the \ion{Ca}{ii} emission is still debated because the decretion disks are believed to be too hot for their occurrence \citep[see][]{shokry2018}. While star clusters offer the possibility to study the origin of \ion{Ca}{ii} emission in a controlled environment with well-defined stellar ages and masses, this endeavour is beyond the scope of our study. However, we emphasise that any line emission will bias our $V\sin i$ measurements to high values. Hence, we refrain from interpreting the $V\sin i$ results of the Be sample. More spectroscopic data, covering metal or helium lines in the blue part of the optical range, are needed.

\subsection{A closer look at the red main sequence}
\label{subsec:red_main_sequence}

\begin{figure}
    \centering
    \includegraphics{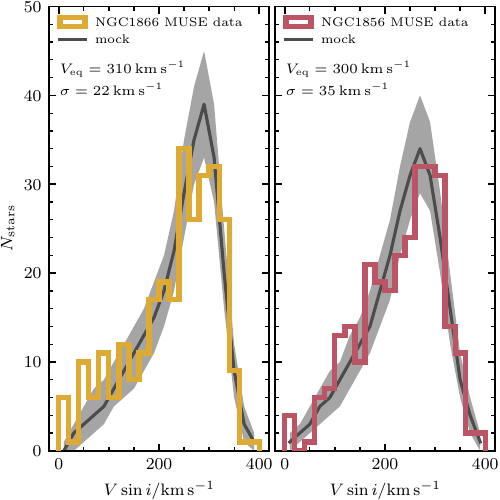}
    \caption{The distributions of $V\sin i$ values obtained for red main-sequence stars in the magnitude range $19<m_{\rm F438W}<20$ in NGC~1866 \textit{(left)} and NGC~1856 \textit{(right)} are shown as solid histograms. The grey lines show the $V\sin i$ distributions expected from the simple model for the equatorial velocity distribution described in the text, with the grey shaded areas indicating the $1\sigma$ confidence intervals for randomly drawing the same number of stars as observed from the models. The model parameter adopted are provided near the top of each panel. In either cluster, the mock distribution provides a reasonable representation of the measured one.}
    \label{fig:vsini_red_ms}
\end{figure}

Although stellar rotation is accepted to play a substantial role in shaping the CMDs of young star clusters, there is still no consensus regarding the actual spin velocities of the stars. Photometric studies based on comparisons between the observed colour spreads and stellar evolutionary models often find that rotational velocities close to the critical value are required for the red main-sequence stars to align model predictions and data \citep[e.g., ][]{milone2017,cordoni2024}. On the other hand, \citet{wang2023} argued that moderately fast rotating stars, spinning at about 60-70\% of the critical value, are sufficient to explain the observed colour spreads and highlighted some discrepancies between the various evolutionary models available. The rotation rates inferred by \citet{wang2023} are in agreement with the direct measurements of $V\sin i$ available for red main-sequence stars in young Magellanic Cloud clusters \citep[e.g.,][]{marino2018,kamann2023}.

To investigate the properties and the evolution of rotational velocities among red main-sequence stars, we focus our attention to the magnitude range $19<m_{\rm F438W}<20$, in which both clusters display a split main sequence (cf. Fig.~\ref{fig:cmd}) and large numbers of $V\sin i$ measurements are available. To select red main-sequence stars, we applied an additional cut of $0.4<\Delta_{\rm F336W,\,F438W}< 1.0$ to both subsamples, resulting in 286 stars in NGC~1866 and 250 stars in NGC~1856. Their $V\sin i$ distributions are shown in Fig.~\ref{fig:vsini_red_ms}. Both distributions peak slightly below $300\,{\rm km\,s^{-1}}$ and are asymmetric, showing an extended tail towards lower values. The latter is expected for an isotropic distribution of inclination angles $i$.

\begin{table}
	\centering
	\caption{Best-fit parameters from the modelling of the observed  $V\sin i$ distribution of red main-sequence stars.}
	\label{tab:model_parameters}
	\begin{tabular}{lcc} 
		\hline
		Cluster & $\langle V_{\rm eq}\rangle/{\rm km\,s^{-1}}$ & $\sigma_{\rm V}/{\rm km\,s^{-1}}$ \\
            \hline
            NGC~1866 & $302.5^{+4.5}_{-4.7}$ & $27.9^{+5.1}_{-5.2}$ \\
            NGC~1856 & $291.3^{+6.0}_{-6.0}$ & $40.5^{+5.6}_{-4.4}$ \\
		\hline
	\end{tabular}
\end{table}

We designed a simple model to recover the intrinsic distribution of equatorial velocities $V_{\rm eq}$ of red main-sequence stars from the observed distributions shown in Fig.~\ref{fig:vsini_red_ms}. We assumed that the intrinsic distribution is Gaussian, with a mean value of $\langle V_{\rm eq}\rangle$ and a dispersion $\sigma_{\rm V}$. Then, the likelihood to observe a star with a projected velocity of $V_{\rm P} \pm \epsilon_{\rm V_P}$ can be approximated as follows.
\begin{multline}
    p(V_{\rm P}, \epsilon_{V_{\rm P}}|\langle V_{\rm eq}\rangle, \sigma_{\rm V}, i)=\\\frac{1}{\sqrt{2\pi\left(\epsilon_{V_{\rm P}}^2 + \sigma_{\rm V}^2 \right)}}\exp{\left[ -\frac{(V_{\rm P} - \langle V_{\rm eq}\rangle\sin i)^2}{2\left(\epsilon_{V_{\rm P}}^2 + \sigma^2 \right)} \right]}.
\end{multline}
The inclination angle $i$ is a nuisance parameter in our case, and cannot be constrained on a star-by-star basis. Hence, we integrate the likelihood over all possible values of $i$, keeping in mind that for an isotropic distribution of spin axes, the probability density function is given by $f(i)=\sin i$.
\begin{multline}
    p(V_{\rm P}, \epsilon_{V_{\rm P}}|\langle V_{\rm eq}\rangle,\, \sigma_{\rm V})=\\\int_{0}^{\pi/2}\frac{1}{\sqrt{2\pi\left(\epsilon_{V_{\rm P}}^2 + \sigma_{\rm V}^2 \right)}}\exp{\left[ -\frac{(V_{\rm p} - \langle V_{\rm eq}\rangle\sin i)^2}{2\left(\epsilon_{V_{\rm P}}^2 + \sigma_{\rm V}^2 \right)} \right]}\sin i\;{\rm d}i.
\end{multline}
The integral can be evaluated numerically for a given set of model parameters ($\langle V_{\rm eq}\rangle,\,\sigma_{\rm V}$) and a measured velocity $V_{\rm P} \pm \epsilon_{\rm V_P}$. We can then determine the likelihood of the model parameters given the data by multiplying the likelihood functions of the individual stars. In practice, we evaluate the $log$-likelihood and explore the parameter space of possible $\langle V_{\rm eq}\rangle-\sigma_{\rm V}$ combinations using \textsc{emcee} \citep{emcee}. The median values and their confidence intervals are listed in Table~\ref{tab:model_parameters}. In both clusters, the red main-sequence stars have mean equatorial velocities of $\langle V_{\rm eq}\rangle\sim300$~\kms. 

To get an idea of how well our simple model reproduces the observed, we randomly draw mock $V_{\rm eq}$ samples with the same number of stars as in the observed samples, project them along the line of sight assuming isotropy, and apply our measurement uncertainties. For each cluster, 1\,000 mock samples are generated and their median values as a function of \vsini{} are shown in Fig.~\ref{fig:vsini_red_ms} as black lines. They provide reasonable representations of the data. However, some differences can be noted. In the case of NGC~1866, our model slightly underestimates the number of $V\sin i$ values below $\sim 100\,{\rm km\,s^{-1}}$, suggesting a small fraction of genuinely slowly rotating stars. We speculate that this population could be composed of photometric binaries that ``migrated'' from the blue main sequence. In the case of NGC~1856, the observed peak in $V\sin i$ appears broader than the one predicted by the model. This is also reflected by the larger value of $\sigma_{\rm V}$ determined for this cluster (cf. Table~\ref{tab:model_parameters}). Interestingly, for this cluster the photometric split between the blue and red main sequences is also less sharp than for NGC~1866, as can be verified in Fig.~\ref{fig:cmd}, even after correcting for differential extinction. This could imply that the intrinsic distribution of rotational velocities is broader for NGC~1856 than it is for NGC~1866. 

\subsection{Comparison to models}

\begin{figure}
    \centering
    \includegraphics{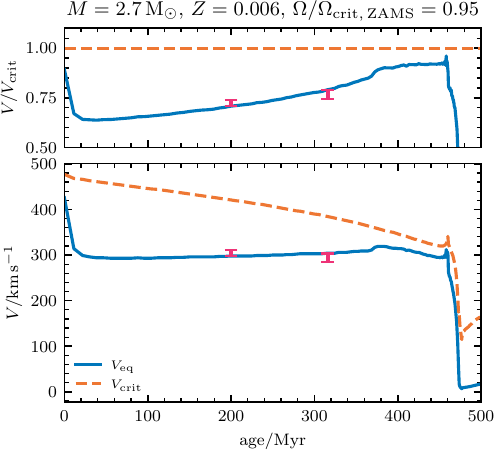}
    \caption{The evolution of the equatorial velocity $V_{\rm eq}$ (solid blue line) and the critical velocity $V_{\rm crit}$ (orange dashed line) as a function of main-sequence lifetime, as predicted by SYCLIST for a $2.7\,{\rm M_\odot}$ star of metallicity $Z=0.006$ and with an initial angular velocity ratio of $\Omega/\Omega_{\rm crit}=0.95$. The top panel shows the same quantities normalized by the critical velocity. In both panels, the mean equatorial velocities derived in this work for NGC~1866 and NGC~1856 are overplotted as the respective ages of the clusters.}
    \label{fig:geneva_models}
\end{figure}

The typical mass of the red main-sequence stars discussed above is $\sim2.7\,{\rm M_\odot}$. In Fig.~\ref{fig:geneva_models}, we compare our measurements of $\langle V_{\rm eq}\rangle$ to the predicted evolution of such stars, based on the SYCLIST models \citep{georgy2013}. The model prediction was calculated assuming a constant metallicity of $Z=0.006$, in agreement with the isochrone fitting presented in Sec.~\ref{subsec:analysis}. We further adopted the same ages ($\sim 200\,{\rm Myr}$ for NGC~1866 and $\sim320\,{\rm Myr}$ for NGC~1856) that provided the best isochrone fits in order to overplot our measurements on the model predictions. Fig.~\ref{fig:geneva_models} shows that the models do not predict any significant evolution of $\langle V_{\rm eq}\rangle$ between the ages of the two clusters, in agreement with our analysis, indicating that no substantial torques brake the stars as they evolve on the main sequence.

To match our results, we need to calculate the models with an initial (zero-age main sequence, ZAMS) angular rotation rate of 95\% critical $(\Omega/\Omega_{\rm crit,\,ZAMS})=0.95$. In terms of the equatorial velocity, this corresponds to an initial ratio of $(V_{\rm eq}/V_{\rm eq,\,crit})\sim0.9$. However, at the current ages of the two clusters, the ratios have dropped to values around $0.7-0.8$. We note that this decrease is mainly caused by an initial drop of the equatorial velocity by $\sim100\,{\rm km\,s^{-1}}$ during the first $\sim20\,{\rm Myr}$ of evolution (see Fig.~\ref{fig:geneva_models}). As discussed by \citet{wang2023}, this drop, which is related to an early relaxation phase of the star, is not seen in other model suites. Furthermore, different models use different definitions of $\Omega_{\rm crit}$ and $V_{\rm eq,\,crit}$. Consequently, care must be taken when inferring fractional critical velocities from measurements. This can be illustrated by the example of NGC~1866, which was included in the works of \citet{gossage2019} and \citet{wang2023}. Both studies compared the cluster's HST photometry to rotating MESA models and found that, to reproduce the main-sequence split, it was sufficient to adopt moderate $\Omega/\Omega_{\rm crit,\,ZAMS}$ or $V_{\rm eq}/V_{\rm eq,\,crit}$ values of around 0.5 for the red main-sequence stars. \citet{wang2023} further showed (their Fig.~1) that to achieve similar colour spreads in the SYCLIST models, higher values of $(\Omega/\Omega_{\rm crit,\,ZAMS})$ between 0.9 and 0.95 were required. This agrees with our findings, depicted in Fig.~\ref{fig:geneva_models}. Indeed, the physical velocities predicted by \citet[their Fig.~10]{gossage2019} across the split main sequence seem to agree well with our measurements.

Very recently, \citet{ettorre2025} performed PARSEC \citep{bressan2012} isochrone modelling of NGC~1866 and argued for a negligible fraction of moderate or slow rotators ($\Omega/\Omega_{\rm crit,\,ZAMS}\lesssim0.8$) in the cluster, in conflict with our direct spin measurements and the studies of \citet{gossage2019} and \citet{wang2023}. Such discrepancies suggest that stellar evolutionary models are not sufficiently accurate to infer stellar spin distributions purely photometrically. In fact, it is extremely challenging to fully reproduce the complex CMDs of YMCs, and the shortcomings of the models can be misinterpreted as age spreads \citep[see discussion in][]{gossage2019} or result in poorly constrained cluster parameters. According to Fig.~13 of \citet{ettorre2025}, their best fit solution for the cluster was found at the high-metallicity end (Z=0.009) of the parameter grid, suggesting that the fit was overly constrained. We suggest that future studies that infer the spin distribution of young clusters through CMD fits should directly incorporate the growing number of \vsini{} measurements when fitting isochrones.

For the remainder of this paper, we will rely on the SYCLIST framework to convert our \vsini{} measurements into critical rotation fractions. To translate our fractional values to other model frameworks, the interested reader is referred to \citet{wang2023}.

A general prediction of stellar evolutionary models is that the critical velocity should decrease as a star reaches the end of its main-sequence lifetime, because its core contracts while its envelope expands. This effect is visualised by the dashed orange curve included in Fig.~\ref{fig:geneva_models}. This drop implies that despite having the same physical equatorial velocities on average, the red main-sequence stars in the older cluster NGC~1856 rotate closer to their critical velocity than those in NGC~1866 by about 10\%. In this respect, it is interesting to note that in the magnitude range investigated, NGC~1856 possesses a significant number of Be stars, while NGC~1866 has none. In both clusters, the red main-sequence stars are expected to be rotating within 20\% of their critical velocities just before leaving the main sequence. We will expand on these aspects in Sec.~\ref{sec:discussion} below when we discuss the origin of the Be stars in the two clusters.

\section{Discussion}
\label{sec:discussion}

\subsection{Stellar rotation}
In both clusters, we measure mean equatorial velocities of $V_{\rm eq}\sim$300~\kms for the red main-sequence stars (cf.~Table~\ref{tab:model_parameters}). The other two young LMC clusters with such measurements available are NGC~1818 and NGC~1850.  For the former, \citet{marino2018} report a mean projected velocity of $\langle$\vsini$\rangle=202\pm23$~\kms{}. Assuming an isotropic distribution of stellar spins, i.e. $\langle \sin i\rangle=\pi/4$, this value translates to a mean equatorial velocity of $\langle V_{\rm eq}\rangle~\sim257$~\kms. For NGC~1850, \citet{kamann2023} found mean equatorial velocities between 230~\kms{} and 280~\kms, depending on the adopted magnitude and analysis method. In comparison to these two studies, our mean velocities are slightly higher, but still broadly consistent, especially when considering the intrinsic dispersion of $V_{\rm eq}$ within the red main-sequence populations of NGC~1866 and NGC~1856 that is visible in Fig.~\ref{fig:vsini_red_ms}.

Turning our attention to the blue main-sequence stars, our analysis confirms that they are relatively slowly rotating, in agreement with model predictions. In addition, Figs~\ref{fig:cmd_vsini} and \ref{fig:vsini_vs_colour} suggest they cover a considerable range of equatorial velocities. Therefore, care must be taken when comparing measurements across different studies and clusters, as the agreement will vary depending on which stars blueward of the red main sequence are included. If we classify as blue main sequence stars all those with $\Delta_{\rm F336W,\,F438W}<0.4$ and again focus on the magnitude range $19<m_{\rm F438W}< 20$, we obtain mean values of $\langle$\vsini$\rangle=91.1\pm10.4$~\kms{} and $\langle$\vsini$\rangle=106.7\pm8.4$~\kms{} for NGC~1866 and NGC~1856, respectively. These values are in agreement with the results for the blue main-sequence stars measured in NGC~1850 \citep[$\langle$\vsini$\rangle=99\pm5$~\kms{}][]{kamann2023} and slightly higher than those of the blue main sequence stars in NGC~1818 \citep[$\langle$\vsini$\rangle=71\pm10$~\kms{}][]{marino2018}.

Another interesting comparison concerns open clusters in the Milky Way, which could help to understand if there are any differences in the spin properties of stars depending on the environment. Many of the Galactic low-mass open clusters do not show split main sequences. In the recent study of \citet{cordoni2024}, two out of 32 clusters do so, NGC~2287 ($\sim$300~Myr) and NGC~3532 ($\sim 410$~Myr). Both clusters have bimodal \vsini{} distributions, with the slow rotators peaking between 75 and 100~\kms and the fast rotators peaking around 250~\kms, in good agreement with the values typically found in the massive Magellanic Cloud clusters.

Considering all spectroscopic measurements made so far, a consistent picture emerges where blue main sequence stars occupy a broad range of equatorial velocities centred around $V_{\rm eq}\lesssim$120~\kms, while fast rotators peak at around 300~\kms in the stellar mass range characteristic for split main sequences (around 3~\msun). Comparison to the critical velocities of the SYCLIST models (cf. Fig.~\ref{fig:geneva_models}) suggests that the blue main sequence stars generally have low $V/V_{\rm crit}\lesssim0.3$ and that fast rotators cover a range of $V/V_{\rm crit}=0.6-0.8$ in the age range investigated so far ($\lesssim300$~Myr). However, the models further predict that the majority of red-main sequence stars will approach their break-up velocities just before evolving off the main sequence (cf. Fig~\ref{fig:geneva_models}). This has implications for the high fractions of Be stars observed in these clusters, as further discussed below in Sec.~\ref{sec:be_fraction}.

As already discussed in Sec.~\ref{subsec:red_main_sequence}, there is no consensus on the initial rotation rates of blue and red main-sequence stars. Studies based on comparisons between the observed photometric colour spreads and those predicted by rotating stellar models yield ambiguous results, with some arguing that initial rotation rates close to critical ($\gtrsim$90\%) are required to reproduce the observed photometric spreads, while others infer moderate values of 50-60\% critical. An in-depth discussion of the model discrepancies is beyond the scope of our work \citep[e.g., see][instead]{wang2023}. However, we argue that in light of such discrepancies, the provision of just the initial rotation rates that reproduce an observation can be misleading, because different initial conditions may result in the same observables. In light of the increasing data base of direct \vsini{} measurements, an effort should be made to consider these measurements when fitting isochrones \citep[e.g.][]{lipatov2022}.

\subsection{The origin of fast and slow rotators}

\begin{figure*}
    \includegraphics{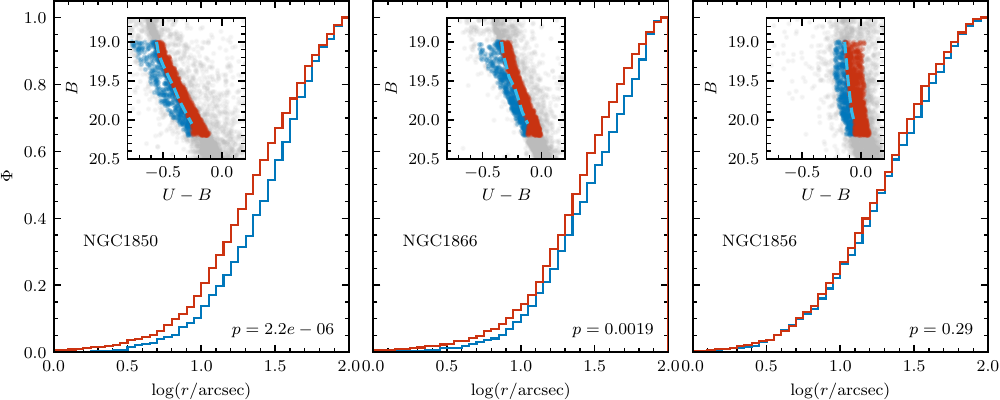}
    \caption{Cumulative radial distributions of red and blue main-sequence stars in NGC~1850 (left), NGC~1866 (centre), and NGC~1856 (right). Note that the clusters are sorted by increasing age. The $p$-value of a two-sided KS test to investigate the similarity of the distributions is provided in the bottom right of each panel. Inset panels illustrate how the two samples were selected from the CMDs of the clusters. Note that for the sake of brevity, we label $m_{\rm F336W}$ and $m_{\rm F438W}$ as $U$ and $B$, respectively.}
    \label{fig:radial}
\end{figure*}

Of the different scenarios that have been proposed to explain the stellar spin distributions inside young star clusters, none makes explicit predictions about the equatorial velocities expected in a given target. Hence, it is difficult to constrain the scenarios based on the \vsini{} measurements presented in Sec.~\ref{sec:rotation}. One aspect that is worth discussing is that while our observations confirm a clear difference between the spin velocities of red and blue main-sequence stars, the bimodality in \vsini{} is less pronounced than one might expect (cf. Fig.~\ref{fig:vsini_vs_colour}). In the scenario proposed by \citet{wang2022}, where blue main-sequence stars are formed via stellar mergers during the first few Myr of cluster evolution, a significant number of intermediate rotators seems difficult to explain. If mergers happen early and result in slowly rotating stars, one would naively expect a well defined peak of low \vsini{} values at the ages of NGC~1856 and NGC~1866. If instead the slow rotators are caused by tidal interactions in binary stars as advocated by \citet{dantona2015}, a range of low equatorial velocities could correspond to a range in binary periods. We note that the average $V_{\rm eq}$ values derived for the blue main sequence stars are comparable to those found for late B-type binaries in the field by \citet{abt2004}. Finally, if stellar rotation is linked to the lifetimes of protostellar disks as proposed by \citet{bastian2020}, a range of equatorial rotation velocities seems plausible. As the final spin of a star depends on exactly when a star loses its disk \citep[see Fig.~2 in][]{bastian2020}, a strong bimodality would be harder to explain instead, as it would require a mechanism that synchronises disk evaporation across a cluster.

\citet{bastian2020} suggested to compare the spatial concentrations of the slow and fast rotators as an observational test of their scenario. As stars born close to the centres of their host clusters are more likely to lose their disks either through photo-evaporation from nearby O-stars or dynamical encounters, their scenario predicts the fast rotators to form centrally concentrated. In the other two scenarios, no such link is immediately expected. We studied the spatial distribution of the slow and fast rotators in NGC~1850, NGC~1866, and NGC~1856 by selecting the stars in the magnitude range $19.0<m_{\rm F438W}<20.2$ and considering as blue (red) main-sequence stars those with pseudo-colours below (above) the 30th percentile of the $\Delta_{\rm F336W,\,F438W}$ distribution (cf. Fig.~\ref{fig:cmd}). The result is shown in Fig.~\ref{fig:radial}, with the inset axes illustrating our selections of blue and red main-sequence stars in the cluster CMDs. We note that assuming a fixed number ratio leads to some cross-contamination between the two groups. This will effectively dampen any signatures of concentration differences, such that our results should be interpreted as conservative estimates of the true segregation between red and blue main-sequence stars.

Interestingly, we observe hints of an age evolution in Fig.~\ref{fig:radial}, in the sense that the youngest cluster NGC~1850 has the most segregated populations and the highest concentration of fast rotators, whereas the oldest cluster NGC~1856 shows fully mixed populations, with NGC~1866 still showing (weaker) signs of segregation. Our results agree well with literature studies available for the individual clusters, namely \citet{Niederhofer2024} for NGC~1850, \citet{milone2017} for NGC~1866, and \citet{li2017} for NGC~1856. One possibility to explain this finding is that the fast rotating red main-sequence stars indeed form centrally concentrated, but that the initial differences are erased as the cluster evolves. However, all clusters should be dynamically young. Assuming a typical scale radius of 25~arcsec and a central dispersion of 4~\kms{} \citep{Niederhofer2024} yields relaxation times around 1~Gyr. Therefore, it seems unlikely that an initial segregation between the two populations would have been completely erased by the age of NGC~1856. We conclude that a larger sample of clusters together with evolutionary cluster simulations \citep[e.g.,][]{wang2023a} are needed in order to confirm or refute the apparent trend in Fig.~\ref{fig:radial}.

\subsection{Be star fraction}
\label{sec:be_fraction}

Numerous photometric and spectroscopic studies of massive LMC/SMC clusters have shown that the fraction of Be stars, \befrac, is a strong function of magnitude, peaking near the MSTO and falling towards fainter magnitudes \citep[e.g.,][]{keller2000, bastian2017, milone2018, bodensteiner2020, kamann2023}.  This trend is also found in the present work for NGC~1866 and NGC~1856.  As such we focus on the shape of \befrac\ as a function of magnitude along with the maximum observed \befrac, \befracmax.

In NGC~1866 and NGC~1856 we find \befracmax$ = 85^{+7}_{-12}\%$ and $78^{+11}_{-17}\%$, respectively. Both NGC~1866 and NGC~1856 have been studied previously, although only NGC~1856 has had its \befrac\ studied in detail \citep[][]{bastian2017, milone2018}.  These studies were done using narrow band H$\alpha$ photometry and found \befracmax$=20-33\%$.  These values are significantly smaller than those found in the present work. 
The likely source of this difference is that photometric studies are less sensitive to H$\alpha$ emission (especially when superimposed on a broad absorption line) than spectroscopy.  At younger ages, when the Be star population is dominated by early-type B stars, the H$\alpha$ emission associated with the Be star phenomenon is observed to be much stronger (see Fig.~\ref{fig:halpha_fluxes}).  Hence, for younger clusters we would expect similar Be star fractions for photometric and spectroscopic studies, but in older clusters photometric studies may significantly underestimate the Be star fraction.
This conclusion is supported by the very young cluster NGC~330 ($\sim$30~Myr) that has had its Be star population studied in detail both photometrically and spectroscopically.  Photometric studies found \befracmax$\sim 55\%$ \citep[][]{milone2018} while spectroscopic estimates were similar, although slightly higher, $68$\% \citep[][]{bodensteiner2020}.

We conclude that \befracmax\ in the massive LMC/SMC clusters studied to date (NGC~330, NGC~1850, NGC~1866 and NGC~1856) is high, from $58$\% to nearly $85$\% in NGC~1866, irrespective of the specific subclass of B-stars present.
These fractions can be used to set constraints on both the origin of Be stars as well as their duty cycle (i.e., how long they can be seen in emission compared to when they are dormant).  These constraints will be discussed in the following sections.

\subsection{The origin of Be stars}
\label{sec:origin_be_stars}

There are two main theories for the origin of Be stars, both of which are likely to occur in nature.  The processes are not exclusive, so it is likely that both models contribute to the formation of Be stars at some level, with their relative importance potentially being a function of stellar mass or environment.  The first model adopts interacting binaries and the second model focusses on the rotational evolution of single stars.  We will briefly outline either model and discuss it in light of our observations.

\subsubsection{Binary Pathway}
\label{sec:binary_model}
The transfer of material from a donor star onto a secondary star can transfer a significant amount of angular momentum, potentially causing the secondary to rotate faster, leading to it reaching the critical rotation rate \citep[see][for a review]{langer2012}.  It is thought that the accretion of just a few percent of a star's total mass may be enough to cause the star to reach critical rotation \citep[][]{packet1981}.  It is typically assumed that the donor star is the primary star of a system, that begins to donate mass onto the secondary once it begins to expand as it leaves the main sequence (case B mass transfer).  Indeed, studies of Be field stars have found evidence for the importance of the binary stars, such as a lack of main-sequence companions and the presence of stripped stars in wide orbits \citep[see][for a recent review]{marchant2024}. As stable mass transfer is assumed to be possible only in binaries with mass ratios close to unity \citep[e.g.][]{wang2020}, Be stars formed through the binary channel are expected to be confined to a narrow magnitude range $\sim$2~mag below the turn-off of a cluster. 

\citet[][]{hastings2021} have modelled a population of Be stars and by intentionally selecting extreme model parameters, have been able to produce an upper limit to the expected \befracmax\ as a function of magnitude.  The most important parameters in this study are 1) the initial binary fraction, 2) the fraction of binaries that will interact and become Be stars (specifically as a function of orbital separation), and 3) the index of the stellar initial mass function (IMF).  Under the assumption of a 100\% binary fraction and that all binary systems will undergo mass-transfer in order to produce a Be star, they find that the upper limit of \befrac\ increases towards to the MSTO, reaching a value of \befracmax$=35\%$ for a top-heavy initial mass function. This value is well below the observed \befracmax in NGC~1866 and NGC~1856.

It is worth exploring the impact of the adopted model parameters.  It is generally accepted that massive stars have high binary fractions, approaching unity with increasing stellar mass \citep{offner2023}.  However, the close binary fraction among B-stars in young massive clusters has been found to be $\lesssim$50\% \citep{banyard2022,saracino2023}.  This appears at odds with the idea that the majority of them gain mass through case B mass transfer.  Even among O-stars, only a minority of the objects accrete mass while on the main sequence \citep{sana2012}.  Taking these considerations into account, the fractions derived by \citet{hastings2021} appear optimistic by a factor of $\gtrsim2$. In that case, the binary channel is unlikely to be the dominant path in the formation of Be stars in the LMC/SMC young massive clusters. This is consistent with the modelling of \citet[][]{vanbever1997} who derived a maximum \befrac\ of $5-20$\% that could come through the binary channel.

\subsubsection{Single Star Path}

The other commonly invoked pathway for Be star formation is through single star evolution.  During hydrogen burning, the critical velocity of a star (needed to form a decretion disk) drops as the core density increases and the envelope expands.  If the equatorial velocity stays constant, the star will evolve closer towards the critical rotation rate as it approaches the MSTO (see Fig.~\ref{fig:geneva_models}).  It is not clear what is the exact fraction of the critical velocity (V$_{\rm eq}/$V$_{\rm crit}$) needed to form a decretion disk, with estimates ranging from 60 to over 95 percent critical rotation (see \S~\ref{sec:rotation_rate_needed}). 

In young massive Magellanic Cloud clusters, the red main-sequence stars make up the majority of the population \citep[$\sim$70\%][]{milone2018} and their rotational distribution typically peaks at $\sim300$~km/s with relatively small dispersion (see Table~\ref{tab:model_parameters}).  Depending on the ages of their host clusters, these rapid rotators are at 60-80\% of their critical rotation (see Fig.~\ref{fig:geneva_models}).  Hence a relatively high fraction of these stars are expected to reach critical rotation as they approach the MSTO. 

\citet{hastings2020} developed a model to predict the Be fraction as a function of age for sets of coeval stars (i.e., clusters).  Their estimate of \befrac\ is based on all stars within one magnitude of the MSTO.  As such, it should be comparable to \befracmax\ as defined in the current work (see Fig.~\ref{fig:be_fraction}).  For the initial rotation distribution, they adopt the measured distribution by \citet[][]{dufton2013}.  We note that this distribution is for late O and early B-stars, which are much more massive than the stars considered in the current work. This distribution also has fewer rapid rotators than found in NGC~1866 and NGC~1856.

The authors find that, depending on the adopted critical rotational threshold, the models predict \befracmax$=30-90\%$ (assuming minimum V$_{\rm rot}/$V$_{\rm \rm crit}$ values from 70-95\%).  Their models predict a sharp increase of the Be-star fraction towards the turn-off, with most Be stars confined to a narrow magnitude range $\sim$0.5~mag. While this shape is not replicated in the observed \befrac\ vs. magnitude distributions shown in Fig.~\ref{fig:be_fraction}, it is likely to depend on the assumed distribution of $V_{\rm eq}$ values.

If we take the split main sequence as an indicator and adopt a typical fraction of red main-sequence stars (fast rotators) of 70\%, assume that most of the fast rotators will become a Be star near the end of their main sequence lifetimes, and further assume that the duty cycle of Be stars is 1, then we might expect 70\% of the B stars in a massive cluster to become Be stars, which is close to the observed fraction. In that respect, it would be insightful to extend the work by \citet{hastings2020} to lower stellar masses and the late B-type stars considered in this work.

\subsubsection{Further observational tests}

One way to constrain the relative importance of single- and binary-star channels in producing Be stars in young massive clusters would be to improve their \vsini{} measurements. In the single-star channel, we would not expect any changes in \vsini{} compared to ordinary B stars, because it is the drop in critical velocity that produces the Be phenomenon. Contrary, in the binary-star channel, the star is spun up by the mass transfer, resulting in elevated \vsini{} values. In that case, NGC~1866 and NGC~1856 provide a unique test case. As can be verified from Fig.~\ref{fig:vsini_vs_colour}, in the magnitude range $19<m_{\rm F438}<20$, NGC~1866 does not show Be stars, while NGC~1856 does. Assuming that these stars started from the same initial values (cf. Fig.~\ref{fig:geneva_models}), we would expect the Be stars in NGC~1856 to be rotating faster than their (non-Be) counterparts in NGC~1866 only if they formed from binary mass transfer. In Fig.~\ref{fig:vsini_vs_colour}, it appears like indeed the Be stars have higher \vsini{} values on average than the normal B stars. However, as discussed in Sec.~\ref{sec:rotation}, we do not consider the \vsini{} measurements as reliable, given the presence of emission features that can bias our spectral analyses of these stars. This is also illustrated by the fact that a significant fraction of the Be stars have fitted \vsini{} values that exceed the critical velocities expected for these stars. Additional observations at higher spectral resolution and with bluer wavelength coverage are needed to settle this case.

Another possibility to weigh the importance of the two scenarios would be binary studies.  Many Be stars in the field have indeed been confirmed to be part of post-interaction binaries \citep[e.g., Be+sdO][]{wang2021}, whereas, as mentioned above, they show a lack of main-sequence companions, in agreement with the expectations from the binary-star channel. Mass transfer will widen the binary orbit, which, in combination with the reduced mass of the companion, will result in a binary that is difficult to detect using line-of-sight velocity variations. In that case, the lower fractions of close binaries detected spectroscopically among Be stars in young massive clusters \citep{bodensteiner2021,saracino2023} could be interpreted in favour of the binary-star channel. Alternatively, however, one could argue that the presence of a close companion slows the rotation of the B star through tidal interaction \citep{abt2004} before its critical velocity drops sufficiently to ignite the Be phenomenon. Hence, the single-star channel also predicts a lower spectroscopic binary fraction of Be stars. Ultimately, binary studies covering longer baselines at higher spectral resolution than currently feasible will be needed to make progress in this respect.

\subsubsection{The rotation rate needed to form Be stars}
\label{sec:rotation_rate_needed}

While rotation has long been posited as the origin of the Be star phenomenon, there has been disagreement on how rapidly a star needs to rotate in order to form a decretion disk (and subsequently be observed as an emission line star). While \citet[][]{townsend2004} suggested that, when taking into account limb darkening, observations are consistent with stars rotating at or over $95$\% critical rotation, \citet{zorec2016} argued for a much wider velocity range, extending down to 30\% critical.  Conversely, \citet[][]{huang2010} found that it could be as low as $60$\%, and also suggested that there could be a stellar mass dependence, with lower mass B stars (like those studied here) requiring $>95$\% of critical rotation.

Using the SYCLIST model from Sec.~\ref{sec:rotation}, we can convert the parameters listed in Table~\ref{tab:model_parameters} to fractional values of the critical rotation, yielding $\langle V/V_{\rm crit}\rangle=0.72\pm0.07$ for NGC~1866 and $0.76\pm0.11$ for NGC~1856. Plugging these values into Gaussian cumulative distribution functions, we find that $\sim$20\% of the stars in NGC~1856 rotate at $\geq$85\% critical, but only $\sim$4\% in NGC~1866. These values agree well with the Be star fractions provided for the two clusters in Fig~\ref{fig:be_fraction} in the magnitude range considered $(19<m_{\rm F438W}<20)$. This simple estimate suggests that -- under the assumption that the majority of the Be stars originate from single stellar evolution -- our data are consistent with an onset of the Be phenomenon at $V/V_{\rm crit}=0.85$ in late B stars.  

If we follow the model evolution depicted in Fig~\ref{fig:geneva_models} to the end of the main sequence lifetimes of the stars under consideration, we see that in both clusters, the peak of the $V/V_{\rm crit}$ distribution will shift to values $\gtrsim0.85$. This implies that the majority of the red main-sequence stars could evolve into Be stars, explaining the high Be fractions observed in the clusters today at slightly higher stellar masses. 

Our results suggest that, at least in these clusters, the single-star channel can produce the majority of the Be stars if rotation at $\sim$85\% of their critical velocity is sufficient to enter the Be phase. Because the clusters studied in this work are populated by late-type B-stars, whereas the importance of the binary pathway has been largely established for early-type B-stars, we argue that the dominant channel to form Be stars may be mass-dependent \citep[see also][]{rivinius2024}. Additionally, our results imply that the duty cycle (the fraction of time that a star spends being observable as a Be star) is near unity. 

\section{Conclusions}
\label{sec:conclusions}

Our analysis of the two young massive clusters NGC~1866 and NGC~1856 confirms that stellar rotation plays a dominant role in shaping their colour-magnitude diagrams. At fixed magnitude, there is a strong correlation between the colour of a star and its projected rotation velocity \vsini{}, which we traced from the extended main sequence turn-offs to the split main sequences of the clusters (see Fig.~\ref{fig:vsini_vs_colour}). Interestingly, we do not observe a strong bimodality between slow and fast rotators in either cluster, as might be expected based on the photometric evidence for distinct blue and red main sequences. Instead, \vsini{} increases gradually with colour, with the bluest stars at a given magnitude often displaying spin velocities below the detection threshold of the MUSE data \citep[$\lesssim30$~\kms][]{kamann2023}. The dominant population of red main sequence stars is characterized by equatorial velocities centred around $V_{\rm eq}\sim300$~\kms{} in either cluster. A comparison to SYCLIST \citep{georgy2013} stellar models suggests that both clusters were formed with comparable stellar spin distributions and that stars approaching the end of their main-sequence lifetimes will predominantly be rotating close to their break-up velocities, i.e. $V/V_{\rm crit}\geq80\%$.

These findings have implications for the rich populations of Be stars uncovered in both clusters. By searching for evidence of \ion{H}{$\alpha$} line emission, we find that Be stars dominate the eMSTOs of NGC~1866 and NGC~1856 in numbers, with maximum fractions of \befracmax$\geq60\%$. In combination with earlier results obtained for the even younger clusters NGC~330 \citep{bodensteiner2020} and NGC~1850 \citep{kamann2023}, we can conclude that young massive clusters show large Be-star fractions regardless of the spectral subclass (B0-B9) populating the turn-off. We note that our measurement of the Be-star fraction in NGC~1856 is considerably above the value derived by \citet{milone2018}, which we attribute to the enhanced sensitivity of spectroscopy versus narrow-band photometry in detecting emission lines.

The formation of Be stars is still debated in the literature, with case-B mass transfer in binaries and single-stellar evolution considered as the alternatives \citep[see the recent review by][]{marchant2024}. The work by \citet{hastings2020,hastings2021} has shown that either mechanism struggles to produce the large number of Be stars observed in young massive clusters. Our \vsini{} measurements in comparison with the SYCLIST models predict that the majority of red main-sequence stars will be rotating at or above 80\% of their break-up velocity as they approach the turn-off. If this threshold is sufficient to trigger the Be phenomenon (and the duty cycle is high), then the fast rotators could naturally explain the majority of Be stars observed in young massive clusters without invoking spin-up via binary mass transfer.

Finally, what do our results imply for the origin of the wide distributions of stellar spins observed in young star clusters? None of the three scenarios proposed in this respect \citep{dantona2015,bastian2020,wang2022} makes detailed predictions about the $V_{\rm eq}$ distribution of slow and fast rotators in clusters with ages of a few 100~Myr. Hence, it is difficult to draw any firm conclusions. It appears that the gradual, rather than abrupt, increase in \vsini{} with colour could help constrain the models. For example, the scenario in which fast rotators lose their protostellar disks earlier than slow rotators \citep{bastian2020} does not a priori predict a bimodality, i.e. discrete populations of slow and fast rotators. The same scenario also predicts that fast rotators should be centrally concentrated relative to slow rotators, in partial agreement with our findings (cf., Fig.~\ref{fig:radial}). In a follow-up study that is presented in Bastian et al.~, A\&A accepted, we investigate whether the populations of blue stragglers found in the clusters can be used to enhance our understanding about the origin of the stellar spin distributions of young massive clusters.

\section*{Acknowledgements}

We thank Julia Bodensteiner and Greta Ettorre for insightful discussions during the revision process. We thank the anonymous referee for their constructive comments.
Based on observations collected at the European Southern Observatory under ESO programme(s) 0102.D-0270(A), 0104.D-0257(B), and 108.2258.001.
This research is based on observations made with the NASA/ESA Hubble Space Telescope obtained from the Space Telescope Science Institute, which is operated by the Association of Universities for Research in Astronomy, Inc., under NASA contract NAS 5–26555. These observations are associated with programs 13379, 14204, and 16748.
SKA gratefully acknowledges funding from UKRI through a Future Leaders Fellowship (grants MR/T022868/1, MR/Y034147/1).
FN acknowledges funding by DLR grant 50 OR 2216.
SS aknowledges funding from the European Union under the grant ERC-2022-AdG, \textit{"StarDance: the non-canonical evolution of stars in clusters"}, Grant Agreement 101093572, PI: E. Pancino.

\section*{Data Availability}
The combined stellar spectra and the results of our spectral analysis is available at CDS via anonymous ftp to cdsarc.u-strasbg.fr (130.79.128.5) or \url{https://cdsarc.cds.unistra.fr/viz-bin/cat/J/MNRAS}. The photometry underlying most of our work has already been published by \citet{Niederhofer2024}. The raw MUSE data are available via the ESO archive. All  auxiliary data will be shared by the authors upon reasonable request.
 



\bibliographystyle{mnras}
\bibliography{references} 






\bsp	
\label{lastpage}
\end{document}